\documentclass[aps,prl,showpacs,reprint]{revtex4-1}

\usepackage{bm}
\usepackage{amsmath}
\usepackage{graphicx}
\usepackage{subfigure}
\usepackage{color}
\renewcommand{\vec}{\bm}
\newcommand{\mleft}{\!\!\!\!\!\! }

\def\ket#1{|#1\rangle}

\begin{document}


\title{Trapped ion emulation of electric dipole moment of neutral relativistic particles}


\author{Tihomir G. Tenev}
\affiliation{Department of Physics, Sofia University, 5 James Bourchier Blvd, Sofia 1164, Bulgaria}
\author{Peter A. Ivanov}
\affiliation{Department of Physics, Sofia University, 5 James Bourchier Blvd, Sofia 1164, Bulgaria}
\author{Nikolay V. Vitanov}
\affiliation{Department of Physics, Sofia University, 5 James Bourchier Blvd, Sofia 1164, Bulgaria}

\date{\today}

\begin{abstract}
The electric dipole moments of various neutral elementary particles, such as neutron, neutrinos, certain hypothetical dark matter particles and others, are predicted to exist by the standard model of high energy physics and various extensions of it.
However, the predicted values are beyond the present experimental capabilities. 
We propose to simulate and emulate the electric dipole moment of neutral relativistic particles and the ensuing effects in the presence of electrostatic field by emulation of an extended Dirac equation in ion traps.
\end{abstract}

\pacs{11.30.Er}

\maketitle

The standard model of particle physics as well as all SUSY models predict that elementary particles should possess an intrinsic electric dipole moment (EDM) aligned with the spin of the particle. There are certain models where dark matter is thought to consist of electrically neutral relativistic particles possessing EDM and/or magnetic dipole moment (MDM)  \cite{Sigurdson}.
The existence of EDM generally requires the breaking of parity and time reversal symmetries.
The validity of the CPT theorem then implies the breaking of CP symmetry.
Mechanisms violating CP symmetry in the standard model are responsible for the generation of EDM of elementary particles with predicted values $d_n\sim 10^{-32}e$ cm for the neutron and $d_e<10^{-33}e$ cm for the electron~\cite{Pospelov}.

The CP violation is also connected with the problem of the observed matter-antimatter asymmetry in the universe.
The SUSY models, which address some of the shortcomings of the standard model, predict much larger values for the EDM of elementary particles:
 $d_e<10^{-26}e$ cm and $10^{-28}<d_n<10^{-25}e$ cm~\cite{Commins}.


The experimental search for the EDM of the neutron, whose neutral charge makes it most suitable for EDM measurement, has been initiated some 60 years ago by Smith, Purcell and Ramsey \cite{Ramsey}.
Decades of effort have also been dedicated to the measurement of the electron EDM. 
A long line of experiments spanning over 60 years 
 have set only upper limits for the EDM's \cite{Hudson,Baker}: $|d_e|<1.05\times10^{-27}e$ cm and $|d_n|<2.9\times10^{-26}e$ cm.

Recently, the field of quantum simulations of physical systems in ion traps has gained momentum including the simulation of relativistic systems.
Proposals have been put forward for simulation of the free Dirac equation \cite{Lamata}, the Dirac oscillator \cite{Bermudez}, the 1D Dirac equation with various Poincare invariant potentials \cite{Casanova}, relativistic Landau levels \cite{RelLandauLev}, and the Majorana equation \cite{MajoranaSim}.
The goal of these simulations have been to experimentally observe analogs of long predicted but never measured effects like Zitterbewegung and Klein's paradox.
Experimental simulation has already been demonstrated for Zitterbewegung in the 1D free Dirac equation \cite{Gerritsma}, and the Klein paradox in a linear potential~\cite{RelKleinExp}.

In this Letter, we propose the simulation of neutral relativistic particle with EDM by means of the Dirac equation.
This allows to experimentally emulate the behavior of EDM in an electrostatic field and two ensuing effects, which have not been measured so far:
 (i) lifting of spin degeneracy by an electrostatic field, and
 (ii) Larmor-like precession of a particle spin in an electrostatic field.
In addition we explore a few relativistic properties of these effects and propose their emulation with trapped ions. 

The Dirac Hamiltonian for a neutral particle with electric dipole and magnetic moments in an external electromagnetic field is given by \cite{Thaller}
\begin{eqnarray}\label{H}
\hat{H} &=& c\hat{\vec{\alpha}}\!\cdot\!\hat{\vec{p}} + \hat{\beta} m c^2 + d_a (\text{i}\hat{\beta}\hat{\vec{\alpha}}\!\cdot\!\vec{B}c + 2\hat{\beta}\hat{\vec{S}}\!\cdot\!\vec{E}) + \nonumber \\ &+& \mu_a(\text{i}\hat{\beta}\hat{\vec{\alpha}}\!\cdot\!\vec{E}/c - 2\hat{\beta}\hat{\vec{S}}\!\cdot\!\vec{B})\; ,
\end{eqnarray}
where $c$ is the speed of light, $m$ is the particle mass, $\hat{\vec{\alpha}}$ and $\hat{\beta}$ are the Dirac matrices, $\hat{\vec{S}}=-\frac{\text{i}}{4}\hat{\vec{\alpha}}\times\hat{\vec{\alpha}}$ is the spin vector operator in relativistic theory, $\vec{E}$ is the electric field, $\vec{B}$ is the magnetic field, $d_a$ is the electric dipole moment of the relativistic particle and $\mu_a$ its magnetic dipole moment.
The extended Dirac Hamiltonian (\ref{H}) is written in its standard form in the lab reference frame.
For $d_a\neq 0$, the Hamiltonian is not invariant with respect to space inversion and time reversal \cite{Thaller}.
For a neutral particle, such as neutron, neutrinos or some hypothetical dark matter particles~\cite{Sigurdson}, moving in a constant electrostatic field with respect to the lab reference frame we have \mbox{$\vec{B}=0$},
 which simplifies the Hamiltonian to the block matrix form
\begin{equation}
\hat{H}=\left[\begin{array}{ccc} mc^2\hat{1} + d_a\hat{\vec{\sigma}}\!\cdot\!\vec{E} &  & c\hat{\vec{\sigma}}\!\cdot\!\hat{\vec{p}} + \text{i}\mu_a\hat{\vec{\sigma}}\!\cdot\!\vec{E}/c \\ c\hat{\vec{\sigma}}\!\cdot\!\hat{\vec{p}} - \text{i}\mu_a\hat{\vec{\sigma}}\!\cdot\!\vec{E}/c &  & -mc^2\hat{1} - d_a\hat{\vec{\sigma}}\!\cdot\!\vec{E}\end{array}\right] ,\label{EQ:Ham}
\end{equation}
where $\hat{\vec{\sigma}}$ is a vector of the Pauli matrices and $\hat{1}$ is the $2\times 2$ identity matrix. In this fully relativistic Hamiltonian the term $\text{i}\mu_a\hat{\vec{\sigma}}\!\cdot\!\vec{E}/c$ coupling the magnetic dipole moment $\mu_a$ to the electrostatic field $\vec{E}$ in the lab frame of reference bears some resemblance to the spin-orbit coupling term which appears in the nonrelativistic limit of the Dirac equation. Classically the spin-orbit coupling term is induced from the Joules-Bernoulli equations as proportional to $\vec{B}_{\bot}\!\!\sim\!\!\vec{v}\!\!\times\!\!\vec{E}$, where $\vec{v}$ is the speed of the particle. However the term $\text{i}\mu_a\hat{\vec{\sigma}}\!\cdot\!\vec{E}/c$ from the fully relativistic Hamiltonian is not the spin-orbit coupling term which although being relativistic in nature appears in the nonrelativistic Pauli equation.
The eigenvectors of $\hat{H}$ are sought in the form of plane waves $|\psi_l\rangle=|l(\vec{p})\rangle e^{-\text{i}\vec{p}\cdot\vec{r}/\hbar}$ where $|l(\vec{p})\rangle$ are four-component spinors with eigenvalue $E_l$.
The general solution of the time-dependent Dirac equation $\text{i}\hbar\partial\Psi/\partial t=\hat{H}\Psi$ is given by \mbox{$\left\vert\Psi\right\rangle=\sum_l b_l |l(\vec{p})\rangle e^{-\text{i}(\vec{p}\cdot\vec{r}+E_lt)/\hbar}$} where $b_l$ are complex-valued coefficients.  

Recently, simulation of the 3D Dirac equation in its supersymmetric representation without external potential has been proposed for simulation of Zitterbewegung of a free electron~\cite{Lamata}.
Following this work we propose here how to simulate Eq.~\eqref{EQ:Ham} in the standard representation.
The simulation of the 3D Dirac equation requires a single ion trapped in a Paul trap, in which the ion oscillates in the three spatial directions $x,y,z$ with frequencies $\nu_j$ ($j=x,y,z$).
The Dirac bispinor $|l(\vec{p})\rangle$ is implemented as a linear combination of four internal ion levels $\ket{a}$, $\ket{b}$, $\ket{c}$ and $\ket{d}$, which represent the internal degrees of freedom of the relativistic particle:  $\ket{l(\vec{p})} = u_a|a\rangle+u_b|b\rangle+u_c|c\rangle+u_d|d\rangle$.
The motional degrees of freedom of the simulated particle can be mapped to the ion vibrations 
 using the relations $\hat{p}_j=\text{i}\hbar(\hat{a}_j^{\dagger}-\hat{a}_j)/(2\Delta_j)$ with $j=x,y,z$, where $\hat{a}_j$ and $\hat{a}_j^{\dagger}$ are the phonon creation and annihilation operators, $\Delta_j=\sqrt{\hbar/2 M \nu_j}$ is the spread in position of the ground-state wave function and $M$ is the ion mass\cite{Leibfried}.

The Hamiltonian \eqref{EQ:Ham} 
 can be implemented by simultaneous application of detuned red-sideband (Jaynes-Cummings, JC),
 blue-sideband (anti-Jaynes-Cummings, AJC) and carrier interactions between appropriately chosen pairs of the ion levels $|a\rangle$, $|b\rangle$, $|c\rangle$, $|d\rangle$.
The detuned JC and AJC Hamiltonians read
\begin{subequations}
\begin{align}
\mleft \hat{H}_j^{\text{JC}} &= \hbar\eta_j\widetilde{\Omega}_j(\hat\sigma^{+}\hat{a}_j e^{\text{i}\phi_r} + \hat\sigma^{-}\hat{a}_j^{\dagger}e^{-\text{i}\phi_r}) + \hbar\delta_j\hat\sigma_z, \\
\mleft \hat{H}_j^{\text{AJC}} &= \hbar\eta_j\widetilde{\Omega}_j(\hat\sigma^{+}\hat{a}_j^{+}e^{\text{i}\phi_b}+\hat\sigma^{-}\hat{a}_je^{-\text{i}\phi_b}) + \hbar\delta_j\hat\sigma_z ,
\end{align}
\end{subequations}
where $\phi_r$ and $\phi_b$ are the red- and blue-sideband phases, $\delta_j$ is the detuning, $\widetilde{\Omega}_j$ are the Rabi frequencies, $\hat\sigma^+$ and $\hat\sigma^-$ are the raising and lowering operators between two pairs of internal ion levels and  $\eta_j=k\sqrt{\hbar/2M\nu_j}$ is the Lamb-Dicke parameter, where $k$ is the wavenumber of the driving field~\cite{Leibfried}.
Homogeneity of space requires to set the trap frequencies in the three spatial directions equal to each other $\nu_x=\nu_y=\nu_z$.
This ensures that $\widetilde{\Omega}_j=\widetilde{\Omega}$, $\Delta_j=\Delta$ and $\eta_j=\eta$ for all spatial directions $j$.

In our proposal the mass term $\beta m_0 c^2$ is implemented by the Stark shift parts of two JC and two AJC interactions applied simultaneously on the transitions $|a\rangle\leftrightarrow|d\rangle$ and $|b\rangle\leftrightarrow|c\rangle$, i.e., we have the mapping $\beta m_0 c^2 \rightarrow 2\hbar\delta\hat\sigma_z^{ad} + 2\hbar\delta\hat\sigma_z^{bc}$.
The momentum term $c\vec{\alpha}\cdot\hat{\vec{p}}$ 
 maps to the three terms
$2\eta\Delta\widetilde{\Omega}(\hat\sigma_x^{ad}+\hat\sigma_x^{bc})\hat{p}_x$, $2\eta\Delta\widetilde{\Omega}(\hat\sigma_y^{ad} - \hat\sigma_y^{bc})\hat{p}_y$, $2\eta\Delta\widetilde{\Omega}(\hat\sigma_x^{ac}-\hat\sigma_x^{bd})\hat{p}_z$,
where the superscripts in the Pauli matrices indicate the internal ion levels between which the couplings are applied. 

The term describing the interaction between the EDM and the electrostatic field
\mbox{$2d_a\beta\hat{\vec{S}}\cdot\vec{E}$} in the Dirac Hamiltonian \eqref{H} is implemented by the carrier interaction \mbox{$\hat{H}^{\text{c}(1)}_{j}=\hbar\Omega_{j}^{(1)}(\hat\sigma^+e^{\text{i}\phi}+\hat\sigma^-e^{-\text{i}\phi})$} with Rabi frequency $\Omega_{j}^{(1)}$, using the mapping
$2d_a\beta\hat{\vec{S}}\cdot\vec{E} \rightarrow 2\hbar\Omega^{(1)}[\hat\sigma_x^{ab} - \hat\sigma_x^{cd},\hat\sigma_y^{ab}-\hat\sigma_y^{cd},\hat\sigma_z^{ab} - \hat\sigma_z^{cd}]$.
The term describing the coupling of the electrostatic field to the MDM is implemented with the help of carrier interaction with Rabi frequency $\Omega_{j}^{(2)}$, using the mapping $\text{i}\mu_a\hat{\beta}\hat{\vec{\alpha}}\!\cdot\!\vec{E}/c\rightarrow 2\hbar\Omega^{(2)}[-\hat{\sigma_y^{ad}}-\hat{\sigma_y^{bc}},\hat{\sigma}_x^{bc}-\hat{\sigma}_x^{ad},\hat{\sigma}_y^{bd}-\hat{\sigma}_y^{ac}]$.
The above terms can be implemented by setting appropriately the phases of the carrier interactions.
This 
establishes the following relationship between the parameters of the Hamiltonian (\ref{EQ:Ham}) and the trapped ion:
\begin{subequations}\label{EQ:Rel}
\begin{eqnarray}
&(\mu_a/c) E_j = 2\hbar\Omega_{j}^{(2)} ,\ \ \ \ &  d_a E_j = 2\hbar\Omega_{j}^{(1)}, \\
& c = 2\eta \Delta \widetilde{\Omega} ,\ \ \ \ &  mc^2 = 2\hbar\delta.
\end{eqnarray}
\end{subequations}
While it is possible to implement the Hamiltonian (\ref{EQ:Ham}) in supersymmetric representation this has the drawback of establishing correlation between the sizes of the emulated mass term and the emulated EDM term. 
The presented proposal for emulation of Eq.~\eqref{EQ:Ham} has the advantage of independent experimental control of the emulated mass and the EDM terms. 

The essential features and effects caused by the interaction of EDM with an electrostatic field remain in the 1D limit, which is experimentally more feasible.
Following Greiner \cite{Greiner} we consider the 1D Dirac equation where the state of the relativistic particle is described by four-component spinors.
This allows for clearer separation of the spin degrees of freedom from the negative and positive energy solutions
and the simulation of lifting of spin degeneracy and Larmor-like precession due to the interaction of EDM with an electrostatic field.
The use of the two-component form of the Dirac equation \cite{Lamata} would not allow this.
The 1D limit of Eq.~\eqref{EQ:Ham} is
\begin{equation}
\hat{H}_{1\text{D}} = c\alpha_x\hat{p}_x + \hat{\beta} mc^2 + 2d_a\hat{\beta} \hat{S}_xE_x + \text{i}(\mu_a/c)\hat{\beta}\hat{\alpha}_x E_x\; .\label{EQ:1DLimit}
\end{equation}
Note that the terms $2d_a\hat{\beta}\hat{S}_xE_x$ and $\text{i}(\mu_a/c)\hat{\beta}\hat{\alpha}_xE_x$ are invariant with respect to Lorentz boosts in the $x$-direction and therefore there is no effective magnetic field seen by the particle in its stationary reference frame. This is another point which differentiates the $\text{i}(\mu_a/c)\hat{\beta}\hat{\alpha}_xE_x$ term from the spin-orbit coupling or $\vec{v}\times\vec{E}$ term which will disappear in 1D limit.
Equation \eqref{EQ:1DLimit} maps to the trapped-ion Hamiltonian
\begin{eqnarray}
\hat{H}_{1\text{D}} &=& 2\eta\Delta\widetilde{\Omega}(\hat\sigma_x^{ad} + \hat\sigma_x^{bc})\hat{p}_x + 2\hbar\delta(\hat\sigma_z^{ad} + \hat\sigma_z^{bc}) + \nonumber \\
 &+& 2\hbar\Omega^{(1)}(\hat\sigma_x^{ab} - \hat\sigma_x^{cd}) - 2\hbar\Omega^{(2)}(\hat{\sigma}_y^{ad} + \hat{\sigma}_y^{bc}) .\label{EQ:1DHam}
\end{eqnarray}
The momentum terms and the mass term in Eq.~\eqref{EQ:1DHam} can be implemented by applying simultaneously two pairs of detuned AJC and JC interactions 
on the transitions $|a\rangle\leftrightarrow|d\rangle$ and $|b\rangle\leftrightarrow|c\rangle$ with $\phi_r=3\pi/2$ and $\phi_b=\pi/2$,
\begin{subequations}
\begin{align}
2\eta\widetilde{\Omega}\Delta\hat\sigma_x^{ad}\hat{p}_x + 2\hbar\delta\hat\sigma_z^{ad} &= \hat{H}_x^{\text{JC},ad}  + \hat{H}_x^{\text{AJC},ad}, \\
2\eta\widetilde{\Omega}\Delta\hat\sigma_x^{bc}\hat{p}_x + 2\hbar\delta\hat\sigma_z^{bc} &=  \hat{H}_x^{\text{JC},bc} + \hat{H}_x^{\text{AJC},bc},
\end{align}
\end{subequations}
with $\delta_x=\delta$.
The implementation of the EDM term $2d_a\beta \hat{S}_xE_x$ requires two carrier interactions with Rabi frequency $\Omega^{(1)}$ on the transitions $|a\rangle\leftrightarrow|b\rangle$ and $|c\rangle\leftrightarrow|d\rangle$:
$2\hbar\Omega^{(1)}(\hat\sigma_x^{ab}-\hat\sigma_x^{cd}) = H^{\text{c}(1)}_{ab}(\phi=0) + H^{\text{c}(1)}_{cd}(\phi=\pi)$.
Using independently and simultaneously two more carrier interactions with Rabi frequency $\Omega^{(2)}$ on the transitions $|a\rangle\leftrightarrow|d\rangle$ and $|b\rangle\leftrightarrow|c\rangle$, the MDM term $\text{i}(\mu_a/c)\hat{\beta}\hat{\alpha}_x E_x$ can be implemented:
 $ - 2\hbar\Omega^{(2)}(\hat{\sigma}_y^{ad} + \hat{\sigma}_y^{bc})=H^{\text{c}(2)}_{ad}(\phi=\pi/2)+H^{\text{c}(2)}_{bc}(\phi=\pi/2)$.

The presence of the EDM term $2 d_a\beta\hat{S}_x\!\cdot\! E_x$ in the Hamiltonian \eqref{EQ:Ham} causes two interlinked effects.
First it leads to lifting of spin degeneracy in the spectrum of the Hamiltonian \eqref{EQ:Ham}, which is caused just by the electrostatic field.
This effect is technically similar to the lifting of spin degeneracy by a static magnetic field, which underpins the anomalous Zeeman and Paschen-Back effects in atomic physics~\cite{Messiah}.
This is explained naturally because the electric dipole moment of any particle is aligned with its spin. However conceptually it is different from SU(2) breaking by magnetic field since lifting of spin degeneracy due to magnetic field involves only the breaking of time-reversal invariance, while lifting of spin-degeneracy by EDM -- electric field coupling breaks both time and space reversal symmetries.
An alternative explanation \cite{Kittel} can be sought in the fact that the electric dipole term $2 d_a\beta\hat{S}_x\!\cdot\! E_x$ breaks the space inversion symmetry \cite{Thaller} while preserving the translational invariance in an electrostatic field.
The second effect is a consequence of the first: 
 the precession of EDM and the associated spin around the electrostatic field, similar to the Larmor precession of a particle spin around a static magnetic field.

While these two effects can also be modeled within non-relativistic limit the simulation of the 1D Dirac equation presents the opportunity to study two unusual purely relativistic features of spin-splitting by electrostatic field. The first is the disappearance of spin-splitting when the mass in the Dirac equation tends to zero. The second is the reduction in the size of the spin-splitting caused by the coupling of the electrostatic field to the MDM embodied in the term $\text{i}(\mu_a/c)\hat{\beta}\hat{\alpha}_xE_x$. The mathematical analysis follows.

For a free Dirac Hamiltonian ($E_x=0$) the positive and negative energy eigenvalues \mbox{$E_\pm=\pm\sqrt{c^2p_x^2+m^2c^4}$} are doubly degenerate reflecting the spin degeneracy.
For a nonzero electrostatic field ($E_x\neq0$) there is no degeneracy in the spectrum of the Hamiltonian \eqref{EQ:Ham},
\begin{subequations}
\begin{align}
E_{\pm}^{\uparrow}&=\pm\sqrt{c^2p_x^2 + E_x^2 (\mu_a/c)^2  +(mc^2+E_x d_a)^2} , \label{EQ:Eup} \\
E_{\pm}^{\downarrow}&=\pm\sqrt{c^2p_x^2 + E_x^2 (\mu_a/c)^2 + (mc^2-E_x d_a)^2} .\label{EQ:Edown}
\end{align}
\end{subequations}
The splitting $\Delta E = E_{+}^{\uparrow} - E_{+}^{\downarrow}  = - (E_{-}^{\uparrow}-E_{-}^{\downarrow})$ is
\begin{align}
\Delta E  
\label{EQ:De}
 = \sqrt{c^2p_x^2 + E_x^2(\mu_a/c)^2 +(mc^2+E_x d_a)^2}\notag\\ - \sqrt{c^2p_x^2 + E_x^2(\mu_a/c)^2 + (mc^2-E_x d_a)^2}.
\end{align}
First, in the limit $m=0$ the spin splitting vanishes, $\Delta E=0$, despite the fact that the EDM and electrostatic field are nonzero. This is in stark contrast to nonrelativistic model where the spin-splitting does not depend in any way on the mass term.
Second, the term $\text{i}(\mu_a/c)\hat{\beta}\hat{\alpha}_xE_x$ does not lead to lifting of the degeneracy as $\Delta E=0$ for $d_a=0$ and $\mu_a\neq0$ and \mbox{$E_x\neq0$} by itself.
Estimates 
of the EDM and MDM of several electrically neutral particles show that $mc^2\gg E_xd_a, (\mu_a/c)E_x$.
Expanding Eq.~\eqref{EQ:De} in Taylor series with respect to the small variable $(\mu_a/c)E_x$ and making the approximations $E_{+}^{\uparrow}\sim E_{+}^{\downarrow}\sim mc^2$ we get for the energy splitting to second order $\Delta E\approx2E_xd_a - \frac{(\mu_a)^2 E_x^2}{c^2}\frac{1}{m^2c^4}d_a E_x$. Therefore the effect of the term $\text{i}(\mu_a/c)\hat{\beta}\hat{\alpha}_xE_x$ is to decrease the energy splitting caused by the EDM -- $E_x$ coupling term. This consequence of $\text{i}(\mu_a/c)\hat{\beta}\hat{\alpha}_xE_x$ again stresses its distinction from the spin-orbit coupling term which has the completely different effect to lift spin-degeneracy by itself and thus to increase the size of spin-splitting on top of of this from the EDM-$E_x$ coupling. Furthermore again the mass of the particle $m$ plays role in the spin-splitting by being one of the determinants of its splitting.
The amount of the decrease of $\Delta E$ due to the MDM -- $E_x$ coupling term depends on the ratio $\lambda=\left(\frac{(\mu_a/c) E_x}{mc^2}\right)^2$.
For practically achievable strengths of $E_x$,  $\lambda<10^{-30}$, thus for conventional experimental setups its effect can be neglected and the energy splitting can be written in the form $\Delta E = \sqrt{c^2p_x^2 + (mc^2+E_x d_a)^2}\notag\\ - \sqrt{c^2p_x^2 + (mc^2-E_x d_a)^2}$.
However the effect of  $\text{i}(\mu_a/c)\hat{\beta}\hat{\alpha}_xE_x$ on $\Delta E$ can be emulated and explored in an ion trap setup.

When $\text{i}(\mu_a/c)\hat{\beta}\hat{\alpha}_xE_x$  is neglected, the
eigenspinors corresponding to $E_{\pm}^{\downarrow}$ and $E_{\pm}^{\uparrow}$ are given by
$\ket{\pm\downarrow} = (v_{\pm}^{\downarrow}, -v_{\pm}^{\downarrow}, -1, 1)^T/N_{\pm}^{\downarrow}$ and
$\ket{\pm\uparrow} = (v_{\pm}^{\uparrow}, v_{\pm}^{\uparrow}, 1, 1)^T/N_{\pm}^{\uparrow}$,
with
$N_{\pm}^{\downarrow}= \sqrt{2+2|v_{\pm}^{\downarrow}|^2}$, $N_{\pm}^{\uparrow} = \sqrt{2+2|v_{\pm}^{\uparrow}|^2}$,
$v_{\pm}^{\downarrow} = w^{\downarrow} \pm \sqrt{1+w^{\downarrow 2}}$, and
$v_{\pm}^{\uparrow} = w^{\uparrow} \pm \sqrt{1+w^{\uparrow 2}}$,
where $w^{\downarrow} = (mc^2-E_x d_a)/(cp_x)$ and $w^{\uparrow} = (mc^2+E_x d_a)/(cp_x)$.
The expectation values of the components $\hat{S}_y$ and $\hat{S}_z$ are zero for all four eigenspinors.
The expectation value of $\hat{S}_x$ for the four-component spinors $\ket{\pm\uparrow}$  is $+\frac{1}{2}$ while the one for $\ket{\pm\downarrow}$ is $-\frac{1}{2}$ confirming the interpretation of Eqs.~\eqref{EQ:Eup} and \eqref{EQ:Edown} as lifted spin degeneracy.



Consider an initial state which is a linear combination of positive energy solutions:
 $ |\Psi(t=0)\rangle = e^{-\text{i}p_x x/\hbar} (b_+^{\uparrow}\ket{+\uparrow} + b_+^{\downarrow}\ket{+\downarrow})$.
Then the general solution of the time-dependent problem will not involve negative energy eigenfunctions:
 $\ket{\Psi(t)} = e^{-\text{i}p_x x/\hbar} [b_+^{\uparrow}\ket{+\uparrow} e^{-\frac{\text{i}}{\hbar}E_+^{\uparrow}t}
  + b_+^{\downarrow}\ket{+\downarrow} e^{-\frac{\text{i}}{\hbar}E_+^{\downarrow}t}]$.
The expectation value of the spin component $\hat{S}_j$ with respect to $|\Psi(t)\rangle$ is
\begin{eqnarray}
\langle \hat{S}_j(t)\rangle &=& |b_+^{\uparrow}|^2 \langle\uparrow+|\hat{S}_j|+\uparrow\rangle + |b_+^{\downarrow}|^2 \langle\downarrow+|\hat{S}_j|+\downarrow\rangle + \nonumber \\
&+& 2\mathrm{Re}\left[ (b_+^{\uparrow})^{*} b_{+}^{\downarrow}\langle\uparrow+\vert\hat{S}_j\vert+\downarrow\rangle e^{\text{i}\omega t}\right] .\label{EQ:Scomp}
\end{eqnarray}
The overall behavior of $\langle\hat{\vec{S}}\rangle$ is a precession around $E_x$ with an angular frequency $\omega$,
 similar to the Larmor precession of a spin in an external magnetic field. 



The experimental signature of the lifting of spin degeneracy by an external electrostatic field is the Larmor-like precession of the spin expectation value.
Using the mapping  
between the relativistic particle and the ion trap parameters the simulated precession frequency becomes
\begin{align}
\omega &= 
 2\sqrt{\eta^2\Delta^2\widetilde{\Omega}^2p_x^2/\hbar^2  + (\delta+\Omega^{(1)})^2} \nonumber \\
  &-2\sqrt{\eta^2\Delta^2\widetilde{\Omega}^2p_x^2/\hbar^2  + (\delta-\Omega^{(1)})^2} .
\end{align}




The simulation requires initialization of the trapped ion in state $\ket{\Psi(t=0)} = e^{-\text{i}p_x x/\hbar}  (b_+^{\uparrow}\ket{+\uparrow} + b_+^{\downarrow}\ket{+\downarrow})$.
The construction of the initial state can be done using the same toolbox which is employed for the simulation of the Hamiltonian \eqref{EQ:1DHam} \cite{Gerritsma}.
The system should first be cooled to its ground state.
Then the motional degrees of freedom should be excited representing certain values of the simulated momentum through the relationship $\hat{p}_x=\text{i}\hbar(\hat{a}_x^{\dagger}-\hat{a}_x)/(2\Delta_x)$.
Then using combination of carrier interactions with appropriate timing for the desired values of the parameters $\Omega,\eta,\Delta,\widetilde{\Omega},\delta$
 one can populate the four ionic levels with the required probabilities.

The dynamics of the system can be driven by the application of the described combination of AJC, JC stemming from a single bichromatic source and carrier interactions will simulate the Dirac Hamiltonian \eqref{EQ:1DHam}. Controlling the transitions between four energy levels is not trivial, but certainly feasible. Current experiments with trapped ions involve even more than four levels, for optical pumping, storage, ancillas, ionization etc. In the proposed experimental implementation we need two pairs of JC and AJC laser fields for the implementation of the momentum term, which couples levels $|a\rangle$ and $|d\rangle$, and $|b\rangle$ and $|c\rangle$. One can always use energy levels that form transitions with different carrier frequencies, and/or transitions driven by different polarizations, thereby eliminating the possibility for interference. We note that light shifts are not essential for the required light intensities; in any case they can easily be accounted for and/or compensated. These arguments are valid also for the implementation of the other terms in the Dirac equation.

The dynamics of the system is manifested in a precession of the emulated relativistic particle spin embodied in oscillations of the relative phase between the two eigenspinors $\ket{+\uparrow}$ and $\ket{+\downarrow}$ with frequency $\omega$. The two eigenspinors $\ket{+\uparrow}$ and $\ket{+\downarrow}$ map to the four internal levels of the trapped ion and the emulated dynamics of EDM precession maps to periodic population transfer between the four internal ion levels with frequency $\omega=(E_{+}^{\uparrow}-E_{+}^{\downarrow})/\hbar$. This frequency $\omega$ which is the signature of the emulated dynamics can be measured by standard ion trap technology such as a electron shelving from any one of the internal ion levels.

Supposing a realistic electrostatic field $E_x=10$ MV/cm a neutron with experimentally set upper EDM value of around $d_n\sim 10^{-26} e$ cm will lead to spin-splitting $\Delta E\sim 10^{-19}$ eV corresponding to precession frequency of $\omega=10^{-4}$ Hz; for a neutron with SM predicted EDM value $d_n\sim 10^{-32} e$ cm the corresponding spin-splitting $\Delta E\sim10^{-25}$ eV and $\omega=10^{-10}$ Hz the precession frequency is so small that it would take of the order of 300 years for one full precession of the spin. These values put considerable challenge to present and future conventional experiments.
The emulation of the Dirac equation with the EDM term, Eq.~\eqref{EQ:1DHam}, provides the possibility for emulation of the discussed effects, since they allow for emulated precession frequencies in the range $\omega\sim10-10^7$ Hz.

In conclusion, we have proposed a scheme for simulating the EDM of neutral relativistic particles within Dirac theory in ion traps. We have described the lifting of spin-degeneracy caused by an electrostatic field for a particle possessing EDM,
and the consequent Larmor-like precession of the particle spin. We have predicted a few unusual relativistic features of the considered effects and proposed how they can be emulated in an ion trap. Furthermore this can serve as a stepping stone towards more involved experimental studies of the physics of combined space inversion and time-reversal violation as well as CP-violation.

This work has been supported by the EU COST project IOTA and the Bulgarian NSF grants D002-90/08 and DMU03/107. 

%


\end{document}